\documentclass[showpacs,amsmath,amssymb,nofootinbib,twocolumn]{revtex4-1}

\usepackage{graphicx}
\usepackage{dcolumn}
\usepackage{bm}
\usepackage{epsfig}
\usepackage{color}

\usepackage{hyperref}
\usepackage{amssymb}
\usepackage{amsmath}
\usepackage{verbatim}
\usepackage{mathrsfs}
\usepackage{amsfonts}
\usepackage{latexsym}
\usepackage{epsfig}
\usepackage{color}
\usepackage{graphicx,subfigure}
\usepackage{units}

\begin{document}

\preprint{}

\title{Horndeski dark matter and beyond}

\author{Alberto Diez-Tejedor, Francisco Flores, Gustavo Niz}

\affiliation{Departamento de F\'isica, Divisi\'on de Ciencias e Ingenier\'ias,
Campus Le\'on, Universidad de Guanajuato, Le\'on 37150, M\'exico}


\begin{abstract}
Starting from the Gleyzes-Langlois-Piazza-Vernizzi action, we derive the most general effective theory that is invariant under internal shifts and a $\mathbb{Z}_2$ 
mirror symmetry in the scalar sector. Contrary to what one may think, this model presents a dark matter tracker previous to the dark energy domination. We show that, 
in an empty universe and to linear order in perturbations, the scalar mode clusters in exactly the same way as standard nonrelativistic cold dark matter. This 
also holds for the subsector of the theory where the speed of propagation of gravitational waves equals that of light, in agreement with the recent multimessenger 
observation. However, the inclusion of standard model particles introduces nontrivial couplings of the gravitational scalar mode to baryons, modifying their clustering 
properties. We argue that no arrangement of the parameters of the model can reduce the extra scalar to precisely behave as cold dark matter.
\end{abstract}


\maketitle
\section{Introduction and summary}

There are two main avenues to address the dark matter problem. The most popular one relies on the existence of some elusive, weakly interacting massive cold 
particles, that were originated in the early universe and shape galactic dynamics today~\cite{Profumo2016}. However, some phenomenological correlations such as the
Tully-Fisher~\cite{Tully1977,McGaugh:2000sr} or the Faber-Jackson's~\cite{Faber1976,Sanders2010} show a puzzling link between baryons and dark matter in 
galaxies~\cite{Persic:1995ru,Gentile:2009bw}, and may suggest that we should keep an eye open to the more speculative possibility of a modification of the gravitation laws at 
certain scales~\cite{Famaey:2011kh}. 

Modifying gravity, however, is not often a simple task, since instabilities and/or large departures from laboratory and solar system constraints commonly 
appear~\cite{Clifton:2011jh}. Lovelock theorem~\cite{Lovelock1972} guarantees that general relativity is the only local gravity theory for the spacetime metric in 
four dimensions that satisfies second order equations of motion. As a consequence, in order to modify general relativity one needs to face with the breakdown of at 
least one of the theorem assumptions. In this paper we consider an additional scalar degree of freedom that might possess higher order equations of motion as a 
mediator of the gravitational interaction apart from the standard helicity two graviton. 

The most general scalar-tensor theory of gravity that still preserves second order equations of motion is the so called Horndeski's model~\cite{Horndeski1974,Deffayet2011}.
Within this theory, explicit subsectors have been considered to explain the present day accelerated expansion of the universe~\cite{Gubitosi:2012hu,Gleyzes:2013ooa,Piazza:2013coa,Bloomfield:2012ff,Gleyzes:2014rba},
and to a lesser extent in the context of dark matter. Particular realizations of a dark matter sector include those in $f(R)$ models~\cite{Cembranos2009} and mimetic 
gravity~\cite{Chamseddine:2013kea,Arroja:2015yvd,Sebastiani:2016ras,Casalino:2018tcd}, to mention some.

Along these lines, Bettoni, Colombo and Liberati~\cite{Bettoni:2013zma} first, 
and more recently Rinaldi~\cite{Rinaldi:2016oqp} and Koutsoumbas {\it et al}~\cite{Koutsoumbas:2017fxp}, have identified a particular term in the quintic sector of 
Horndeski gravity that mimics the homogeneous background evolution of a nonrelativistic cold dark matter component. This term emerges from the coupling of the 
Einstein tensor to the gradients of a scalar field, namely $G_{\mu\nu}\nabla^{\mu}\phi\nabla^{\nu}\phi$, which upon integration by parts can be shown to be also 
a part of the quartic Horndeski Lagrangian. In this work we show that the dark matter phenomena is more generic in Horndeski gravity than previously explored, and  
not only this term, but any other one nonminimally coupled to gravity and invariant under internal shift and $\mathbb{Z}_2$ mirror transformations leads to the 
same background evolution.

A distinctive feature of these nonminimally coupled terms in the Horndeski action is that they predict a speed of propagation for the gravitational waves that is, 
in general, different to that of light~\cite{Lombriser:2015sxa,Lombriser:2016yzn,Gubitosi:2012hu,Gleyzes:2013ooa,Piazza:2013coa,Bloomfield:2012ff,Gleyzes:2014rba}. However, the 
recent detection of the gravitational wave signal GW170817 (presumably from a neutron star merger), along with its almost simultaneous gamma ray counterpart 
GRB170817A~\cite{Monitor:2017mdv}, severely constrains any possible deviation in the speed of propagation of the two 
sources~\cite{Ezquiaga:2017ekz,Creminelli:2017sry,Baker:2017hug,Sakstein:2017xjx,Amendola:2017orw}.
As a consequence, we find that a successful model including any of the surviving pieces of the effective Lagrangian should be necessarily a part of an extended version of the Horndeski theory. 
One possibility is to allow higher order derivatives in the equations of motion, but without introducing ghostly or tachyonic propagating modes (see Linder's no slip 
gravity~\cite{Linder:2018jil} for a more conservative proposal, but where the imposed symmetries leading to the dark matter behavior are not present). In this paper, 
we concentrate on the Gleyzes-Langlois-Piazza-Vernizzi scalar-tensor theory~\citep{Gleyzes:2014dya,Gleyzes:2014qga}, a particular realization of a healthy gravity 
theory beyond Horndeski~\cite{Zumalacarregui:2013pma}. As we will find, and after imposing the internal shift and mirror invariance, the new action still respects the same background evolution as 
in the Horndeski model, but with the possibility of a vanishing tensor speed excess.

In order to take seriously the possibility of a cold dark matter component with a gravitational coupling to the standard matter, one needs to explore further the 
full evolution of the scalar mode, and in particular of its perturbations, during the different phases of the universe. We prove that, in an idealized universe empty 
of particles, the new degree of freedom clusters in exactly the same way as a nonrelativistic matter component does. This result is independent of the arbitrary 
functions in the effective Lagrangian. However, the inclusion of standard model particles leads to nontrivial couplings between the scalar field and the visible 
sector, modifying their clustering properties. We argue that there is not any possible choice of the parameters that can accommodate a gravitational mode that mimics 
the behavior of a standard cold dark matter degree of freedom during the matter era, even if only the linear order perturbations are considered. 

The paper is organized as follows: in Section~\ref{sec.theoretical} we describe our starting point, based on the Gleyzes-Langlois-Piazza-Vernizzi action, together 
with the internal shift and $\mathbb{Z}_2$ mirror symmetries of the scalar sector. Then in Section~\ref{sec.DM}, we describe how the homogeneous background evolution 
associated to this modified gravity model resembles that of a cold dark matter component, no matter the form of the arbitrary functions in the Lagrangian. Next, we 
discuss the evolution of linear order perturbations in the case of an empty universe, Section~\ref{sec.EdS}, and also in the presence of matter, 
Section~\ref{sec.DMperturbations}. Finally we conclude with a discussion of the main results in Section~\ref{sec.discussion}. Complementary information can be found 
in the Appendices.

\section{Theoretical framework}\label{sec.theoretical}

Our starting point is the Gleyzes-Langlois-Piazza-Vernizzi (GLPV) scalar-tensor theory~\citep{Gleyzes:2014dya,Gleyzes:2014qga}, described in terms of the following 
action:
\begin{equation}\label{eq:Hornd-action}
S=\int d^4x\sqrt{-g}\left[\sum_{i=2}^5\mathcal{L}_i[g_{\mu\nu},\phi]+\mathcal{L}_m[g_{\mu\nu},\Psi]\right].
\end{equation} 
In this theory the gravitational sector is characterized in terms of 6 possible independent pieces,\footnote{Notice that there are two different contributions to the 
quartic and quintic terms that are not related {\it a priori}.} namely
\begin{eqnarray}
\mathcal{L}_2&=& G_2(\phi , X), \nonumber \\
\mathcal{L}_3 &=& G_3(\phi , X)\square\phi, \nonumber\\
\mathcal{L}_4 &=& G_4(\phi , X)R-2G_{4X}(\phi , X)\left[(\square\phi)^2-\phi_{;\mu\nu}\phi^{;\mu\nu}\right] \nonumber\\
&&+F_4(\phi,X)\epsilon^{\mu\nu\rho}{}_{\sigma}\epsilon^{\mu'\nu'\rho'\sigma}\phi_{;\mu}\phi_{;\mu'}\phi_{;\nu\nu'}\phi_{;\rho\rho'} ,\nonumber\\
\mathcal{L}_5 &=& G_5(\phi , X)G_{\mu\nu}\phi^{;\mu\nu}\nonumber\\
&&+\frac{1}{3}G_{5X}(\phi , X)\left[(\square\phi)^3
+2\phi_{;\mu}^{\nu}\phi_{;\nu}^{\alpha}\phi_{;\alpha}^{\mu}-3\phi_{;\mu\nu}\phi^{;\mu\nu}\square\phi\right] \nonumber\\
&&+F_5(\phi,X)\epsilon^{\mu\nu\rho\sigma}\epsilon^{\mu'\nu'\rho'\sigma'}\phi_{;\mu}\phi_{;\mu'}\phi_{;\nu\nu'}\phi_{;\rho\rho'}\phi_{;\sigma\sigma'}.
\end{eqnarray}
The expression 
\begin{equation}
 X\equiv g^{\mu\nu}\partial_\mu\phi\partial_\nu\phi
\end{equation}
denotes the canonical kinetic term associated to the scalar field, while $R$ and $G_{\mu\nu}$ are 
the standard Ricci scalar and Einstein tensor, respectively, defined with respect to the spacetime metric, $\epsilon_{\mu\nu\rho\sigma}$ is the totally antisymmetric 
Levi-Civita tensor, and semicolons denote covariant derivatives. We choose to work with natural units where $8\pi G=\hbar=c=1$, such that all quantities are dimensionless,
and assume the matter fields $\Psi$ are minimally coupled to the metric only, $\mathcal{L}_m[g_{\mu\nu},\Psi]$, so that the weak equivalence principle is guaranteed. 
 
Although the equations of motion associated to the action~(\ref{eq:Hornd-action}) are of higher order, the true propagating degrees of freedom obey second order 
equations, avoiding Ostrogradski instabilities~\citep{Gleyzes:2014dya,Gleyzes:2014qga,Crisostomi:2016tcp}. At this point, the quantities $G_i(\phi , X)$ and 
$F_i(\phi , X)$ are arbitrary functions of the scalar field and its canonical kinetic term, where the subindex $X$ refers to partial differentiations with respect to 
this variable. Note that if $F_4=F_5=0$, one recovers the more familiar Horndeski model~\cite{Horndeski1974,Deffayet2011,Kobayashi:2011nu}. 

Symmetries are aesthetically appealing mathematical features to physicists, playing a fundamental role in the construction of theories in Nature, such as the standard
model of particles. In this spirit, we explore a sector of the GLPV action that satisfies two symmetries: a shift symmetry 
$\phi\rightarrow \phi + c$, for $c$ an arbitrary constant, and a discrete $\mathbb{Z}_2$ mirror symmetry $\phi \rightarrow -\phi$. The purpose of these symmetries is 
to reduce the GLPV original proposal to a broad sector which admits cosmological dark matter. Among the surviving pieces in the action we find specific constructions 
such as those of Refs.~\cite{Bettoni:2013zma,Rinaldi:2016oqp, Koutsoumbas:2017fxp}, which we analyze now in a systematic way. Notice that there are other proposals in the market 
where a gravitationally coupled scalar degree of freedom plays the role of dark matter and which do not fulfill the above mentioned symmetries, e.g.~in $f(R)$ 
theories~\citep{Cembranos2009}, or in mimetic gravity~\citep{Chamseddine:2013kea,Arroja:2015yvd,Sebastiani:2016ras,Casalino:2018tcd}, but where the desired behavior 
does not seem to arise generically but on a case by case basis. In this context, one may look at the recent work of Ref.~\cite{Langlois:2018jdg}, which shows an embedding of mimetic gravity in the DHOST 
models~\cite{Langlois:2015cwa,BenAchour:2016fzp}, which are in turn a further extension of the GLPV theory.

Some comments about the symmetries of our theory are in order. The shift symmetry keeps terms which only involve derivatives of the scalar field, that may allow a 
restoration of general relativity within the solar system via the Vainshtein mechanism; see for example~\cite{Vainshtein:1972sx,Koyama:2013paa}. Actually, when $F_4$ 
or $F_5$ are included, the Vainshtein mechanism can get suppressed inside matter sources~\cite{Kobayashi:2014ida}, leading to interesting detection perspectives.
The second imposed symmetry plays a not so obvious role, but in practice reduce the number of nonminimally coupled terms in the Lagrangian to those that generically 
present a dark matter behavior. 

When applying these symmetries to the action in Eq.~(\ref{eq:Hornd-action}), $F_5$ vanishes, while $G_2$, $G_4$, and $F_4$ are forced to be functions of the canonical
kinetic term $X$ only. The remaining terms in the Lagrangian,  $G_3$ and $G_5$, do not vanish, but can only be linear in the scalar field $\phi$. However, one may 
show upon integrations by parts that these linear functions of the scalar field in the cubic and quintic terms are equal to linear functions of the kinetic scalar in 
the quadratic and quartic sectors, respectively, so they can be reabsorbed in the second and fourth pieces of the original action. It is important to mention that 
the $G_5$ term left by the symmetries, which can also be written as $G_{\mu\nu}\nabla^\mu \phi\nabla^\nu \phi$, is a popular piece of the Horndeski theory which has 
gained recent attention for dark energy and compact objects, see e.g.~\cite{Chagoya:2016aar,Chagoya:2017fyl}. In the context of dark matter, this coefficient alone 
leads to a negative speed of sound squared during the radiation domination era, as has been previously discussed in Ref.~\cite{Rinaldi:2016oqp}. However, one may 
show that when a more general term of the form $G_4=X^n$ is considered (which reduces to the $G_5=\phi$ for $n=1$), this instability goes away for $n\ge 7/6$. 

In the infrared, and assuming that the arbitrary functions of the Lagrangian are analytical, this model gets dominated by the constant part of $G_2$ (that we assume 
nonzero in this paper), giving rise to the appearance of an effective cosmological constant in the late universe. At higher energies, however, we assume that the 
$X$ dependent coefficients of $G_4$ dominate over those of $G_2$, shaping the evolution of the scalar mode prior to the dark energy domination. Put in a different 
way, since we are particularly interested in the sector where the scalar field is nonminimally coupled to gravity, we will restrict our attention to the case of 
$G_2=\textrm{const}$. In order to be consistent with current cosmological observations, we consider that this constant is determined only by the dark energy physics, 
and fix $G_2=-\Lambda$. 

In conclusion, the sector of the GLPV action~(\ref{eq:Hornd-action}) left after imposing the parity and shift symmetries, and forgetting total derivatives and 
minimally coupled terms, is 
\begin{widetext}
 \begin{equation}\label{eq:Action}
S=\int d^4x \sqrt{-g}\left[\frac{R}{2}-\Lambda+\bar{G}_4(X)R -2\bar{G}_{4X}(X)\left[(\square\phi)^2-\phi_{;\mu\nu}\phi^{;\mu\nu} \right] 
+F_4(X)\epsilon^{\mu\nu\rho}{}_{\sigma}\epsilon^{\mu'\nu'\rho'\sigma}\phi_{;\mu}\phi_{;\mu'}\phi_{;\nu\nu'}\phi_{;\rho\rho'}
+ \mathcal{L}_m \right].
\end{equation}
\end{widetext}
With no loss of generality we have shifted the original $G_4(X)$ function by $G_4(X)= 1/2 + \bar{G}_4(X)$ to explicitly show the Einstein-Hilbert piece, thus 
recovering general relativity plus a cosmological constant $\Lambda$ when $\bar{G}_4=F_4=0$. The action (\ref{eq:Action}) with the extra scalar $\phi$ as a part 
of the gravitational interaction is our starting point to discuss how a dark matter component may naturally arise within some general sectors of the GLPV gravity. 

Before we proceed, however, it is important to place our model within the common belief that Hordenski gravity naturally leads to dark energy scenarios. It is well 
known that shift-symmetric Horndeski theories, like e.g. the covariant Galileon field, may present a de Sitter tracking solution~\cite{DeFelice:2010pv,Germani:2017pwt}. 
Therefore, one may wonder how a dark matter scenario arises when the extra mirror symmetry is considered. From the analysis of 
Refs.~\cite{DeFelice:2010pv,Germani:2017pwt}, however, it is straightforward to get convinced that when one only considers the $G_4$ term, the de Sitter tracker 
conditions cannot be fulfilled, in agreement with our findings. Moreover, this can be generalized to theories that lie beyond Horndeski if one includes an $F_4$ 
function that depends only on the kinetic term~\cite{uslater}

\section{Dark Matter}\label{sec.DM}

The purpose of this section is to convince the reader that dark matter is a generic feature of our general action~(\ref{eq:Action}). In the case of a spatially flat 
Robertson-Walker background, the spacetime metric is described in terms of the line-element
\begin{equation}
 ds^2 =-dt^2+a^2(t)[dx^2+dy^2+dz^2],
\end{equation}
and the contribution of a homogeneous field distribution $\phi=\phi(t)$ to the modified Friedmann equations reads:
\begin{eqnarray}\label{eq:modified-Friedmann}
 3H^2 &=& \rho_m + \rho_\Lambda + \rho_\phi,\\
-2\dot{H}-3H^2 &=& p_m+ p_\Lambda+ p_\phi,
\end{eqnarray}
where ($t,x,y,z$) is a spacetime coordinate system comoving with the expansion, $a$ is the scale factor, 
$H\equiv \dot{a}/a$ is the Hubble parameter, and the overdot represents a derivative with respect to the comoving cosmological time. The expressions 
$\rho_m$ and $p_m$ ($\rho_\Lambda$ and $p_\Lambda$)
denote, respectively, the energy density and pressure of matter, i.e. standard model particles (dark energy, i.e. cosmological constant), while~\cite{Gleyzes:2014rba} 
\begin{eqnarray}
\rho_\phi &=& 3H^2\left[1-M_*^2(1+\alpha_B)\right],\label{eq.rhophi}\\
p_\phi &=& -3H^2\left(1+\frac{2\dot{H}}{3H^2}\right)\left[1-\frac{M_*^2(\alpha_K+6\alpha_B^2)}{\alpha_K}\right]\label{eq.pphi}
\end{eqnarray}
are the corresponding counterparts for the scalar field. For convenience, we have introduced the Bellini and Sawicki's (BS) parametrization of Horndeski 
gravity~\cite{Bellini:2014fua}, consisting on 4 functions of time: $M_*^2$, $\alpha_B$, $\alpha_k$ and $\alpha_T$, extended to include the new term in the GLPV model
via the also time-dependent function $\alpha_H$~\cite{Gleyzes:2014qga} (see Appendix~\ref{functions} for their expressions in 
terms of the original $\bar{G}_4$ and $F_4$ functions in the Lagrangian). 

The conservation equations for matter and the scalar field are the usual, $\dot{\rho_i}+3H(\rho_i+p_i)=0$, which for the latter one yields to 
\begin{eqnarray}\label{eq.dynamics0}
 \dot{X} = 18HX\left(1+\frac{2\dot{H}}{3H^2}\right)\left(\frac{\alpha_B}{\alpha_K}\right).
\end{eqnarray}
Since the theory is invariant under shift transformations, the scalar field does not appear explicitly in the equation of motion. The parameters $M_*^2$, $\alpha_B$ 
and $\alpha_K$ in Eqs.~(\ref{eq.rhophi}), (\ref{eq.pphi}), and~(\ref{eq.dynamics0}) denote the cosmological strength of gravity and the popular braiding and kineticity 
functions, respectively. Note that the other two functions of the BS parametrization, the tensor speed excess, $\alpha_T$, and the coefficient capturing the effects 
beyond Horndeski, $\alpha_H$, do not show up at the level of the homogeneous background.

Before continuing, we would like to make some additional comments on the background equations and the BS functions.
From the expression in Eq.~(\ref{eq.rhophi}) and the definition of the dimensionless density parameter, we can write:
\begin{equation}\label{eq.dimensionless}
 \Omega_{\phi}=1-M_*^2(1+\alpha_B).
\end{equation}
In order to have a significant impact on the evolution of the universe, we can easily conclude that a modification of gravity in the form of its strength, 
$M_*^2\neq 1$, or braiding, $\alpha_B\neq 0$, is necessary. Furthermore, from the perturbation equations that we will discuss later, only tensor modes propagate at 
linear order if $\alpha_B=\alpha_K=\alpha_T=\alpha_H=0$. However, a new degree of freedom apart from those in general relativity and the standard model of particles is 
necessary to explain, e.g., the onset of structure formation during the radiation era, or the baryon acoustic oscillation signal on the large scale structure. We can then 
conclude that something beyond a purely modification of the strength of gravity on cosmological scales, given by $M_*$ in the BS parametrization, must emerge in 
order to have a successful cold dark matter candidate.

The main reason why this model naturally behaves as cold dark matter, at least at the level of the background universe, is because during the matter domination era 
the factor $1+2\dot{H}/{3H^2}$ in Eqs.~(\ref{eq.pphi}) and~(\ref{eq.dynamics0}) naturally vanishes. This results in a frozen kinetic term, $\dot{X}=0$, with a 
vanishing associated pressure, $p_{\phi}=0$. Therefore, during matter domination, the energy density of the scalar field evolves as $\rho_\phi a^3\sim\mathrm{const}$,
as one may appreciate from Eq.~(\ref{eq.rhophi}), contributing as nonrelativistic particles. However, one has to understand the full evolution of the scalar degree 
of freedom and its perturbations during the different domination phases of the universe to really asses whether this model is a viable dark matter candidate. 
 
Before moving to the perturbations, it is interesting to notice a particular realization in which the background scalar field tracks the dominant component not
only during the matter era, but also during the other stages of the evolution. From Eq.~(\ref{eq.dynamics0}), and if the braiding function vanishes, $\alpha_B=0$, 
the kinetic term will remain frozen no matter the value of the other factors in the dynamical equation. Introducing this into Eqs.~(\ref{eq.rhophi}) 
and~(\ref{eq.pphi}), we can easily identify that $p_{\phi}/\rho_{\phi}=-(1+2\dot{H}/3H^2)=w$, where $w$ is the equation of state parameter of the dominant 
matter component. We will comeback to this situation later. 
 
In what follows, we analyze the behavior of the small deviations with respect to the homogeneous and isotropic solution and conclude that, for a universe that is 
empty of matter, linear order perturbations naturally grow like in the standard cold dark matter scenario. Then in Section~\ref{sec.DMperturbations} we consider the 
inclusion of matter.

\section{Empty universe}\label{sec.EdS}

Linear order perturbations are fully characterized in terms of the 5 background functions of the BS parametrization in Appendix~\ref{functions}. In the context of a perturbation theory, these 
functions are related to those of the so called effective field theory (EFT) of dark energy, which unify most of the single degree of freedom dark energy 
models~\cite{Gubitosi:2012hu,Gleyzes:2013ooa,Piazza:2013coa,Bloomfield:2012ff,Gleyzes:2014rba}, including those which are nonminimally coupled to gravity. 
Note that the kinetic scalar remains frozen in this simplified version of the universe, $\dot{X}=0$, and then all these functions become constant parameters, which 
simplifies the analysis. In a universe with no matter, $\Omega_{\phi}=1$, the closure relation in Eq.~(\ref{eq.dimensionless}) fixes the value of the braiding 
coefficient to $\alpha_B=-1$, leaving the other BS parameters arbitrary.  

In general, tensor perturbations are modified with respect to those in general relativity, as can be appreciated from Eq.~(\ref{eq.actiontensor}) in one of the 
Appendices. This expression also codifies the stability conditions of the tensor sector. In order to avoid ghost-like instabilities, one needs $M_*^2>0$. However, 
more interestingly is that the speed of propagation of gravitational waves may differ with respect to that of light for a nonvanishing $\alpha_T$, namely 
$c_g^2=1+\alpha_T$. The almost simultaneous observation of two different signals, one in the form of gravitational waves and the other in gamma rays, coming from a 
same astrophysical event~\cite{Monitor:2017mdv}, constrains the tensor speed excess to $\alpha_T=0$ (with possible deviations from this value smaller than one 
in $10^{15}$)~\cite{Ezquiaga:2017ekz,Creminelli:2017sry,Baker:2017hug,Sakstein:2017xjx,Amendola:2017orw}. Notice that this naturally guarantees the absence of gradient instabilities 
in the tensor sector. 

In addition to tensor perturbations, a scalar mode also propagates in this theory. This mode is expected to host the dark matter sector in this model, and plays a 
central role in our presentation. In the unitary gauge, and integrating out the Hamiltonian and momentum constraints, we obtain the following action for the 
scalar degree of freedom~\cite{Gleyzes:2014dya,Gleyzes:2014qga,DeFelice:2015isa} 
\begin{equation}\label{eq.actionscalargravity}
S_{\textrm{scalar}}^{(2)}=\frac{1}{2}\int dtdx^{3}a^3Q_s\left[\dot{\zeta}^2+(c^0_s)^2\frac{\partial^2\zeta}{a^2}\right],
\end{equation}
where $\zeta$ is the curvature perturbation [$^{(3)}R=-4a^{-2}\partial^2\zeta$], and $Q_s$ and $(c^0_s)^2$ are the kinetic coefficient and speed of sound in vacuum, 
respectively, given by Eqs.~(\ref{eq.kinetic}) and~(\ref{eq.speed}) in the Appendix. From the second of these expressions we can easily identify that $(c_s^0)^2\sim (1+\alpha_B)$. 
This guarantees that the speed of propagation of the scalar mode necessarily vanishes in an empty universe. Moreover, in order to guarantee a theory with no ghosts, 
we need to impose $Q_s>0$, i.e., $\alpha_K>-6$.

Varying the action (\ref{eq.actionscalargravity}) with respect to the curvature perturbation, we obtain 
\begin{equation}
\ddot{\zeta}+3H\dot{\zeta}=0.
\end{equation}
The general solution to this 
equation is a linear combination of a constant term, and a function that decreases with the cosmic expansion as $1/a^3$. Moving to the more familiar Newtonian (or 
longitudinal) gauge, this solution translates into
\begin{equation}\label{eq.ev.perts}
\zeta_{\textrm{Newt.}}= C_1(\vec{x})+\frac{C_2(\vec{x})}{a^5},
\end{equation}
with $C_1$ and $C_2$ two arbitrary integration functions that depend only on the spatial coordinates. This is the well-known behavior of curvature perturbations in 
an Einstein-de Sitter universe (i.e. a universe dominated by nonrelativistic matter), the constant solution being the one associated to the growing mode in the 
density contrast responsible of structure formation. 

Notice that two of the BS parameters are determined in this model, $\alpha_B=-1$ and $\alpha_T=0$, but the other three remain arbitrary. Since the kinetic scalar is 
frozen for this simplified version of the universe, the five BS parameters remain independent: we can just infer the value of the BS functions at a given point $X$, 
which are not related one to the others, but we cannot say anything about their functional dependence; see however the discussion in Section~\ref{sec.DMperturbations}. 
This leaves place for a universe dominated by a cold dark matter degree of freedom where no tensor speed excess is manifest. Let us analyze how this picture is 
modified by the inclusion of standard model particles.

\section{Introducing matter}\label{sec.DMperturbations}

Adding matter to the universe necessarily moves the braiding coefficient away from $\alpha_B= -1$, see Eq.~(\ref{eq.dimensionless}), avoiding to straightforwardly 
guarantee the vanishing of the sound speed of scalar perturbations. This certainly modifies the clustering properties of the extra mode. Furthermore, their 
nontrivial couplings to matter may also change the way in which photons and baryons behave, as has been previously discussed in, e.g., Refs.~\cite{Gleyzes:2014dya,Gleyzes:2014qga,DeFelice:2015isa}. 
Finally, the presence of matter will increase the complexity of the evolution of the background universe with respect to the naive description of Section~\ref{sec.EdS}. All 
these issues deserves a more careful analysis, and we devote this section to this purpose.

At the level of the homogeneous background, the scalar field must be subdominant during radiation domination, scaling with the scale factor accordingly to the particular 
expressions for $\bar{G}_4$ and $F_4$ chosen in the theory. Notice that these functions do not necessarily need to produce a vanishing tensor speed excess $\alpha_T=0$ at that time, since 
the observational multimessenger constraint only requires this to hold on the local late universe. After matter-radiation equality, the scalar mode and the baryons 
take over and drive the cosmological expansion, with the energy density scaling as $1/a^3$ and the canonical kinetic term frozen at a given value. An oversimplified 
description of this period was presented in the previous section. On the more recent epoch, dark energy dynamics prevail. 

The only deviations away from the $\Lambda$CDM model are at the radiation-matter and matter-dark energy transitions, where depending on the particular shape of 
$\bar{G}_4$ and $F_4$ one can get different percent departures from the standard predictions. At the second transition, background observations do not have enough 
precision to discriminate models. In the case of the radiation-matter transition, perturbations are the only door to observations.

Assuming analyticity and no fine tuning, a natural way to satisfy that the modifications to the standard Friedmann equations are 
negligible before matter-radiation equality is by demanding $\bar{G}_4(X=0)=F_4(X=0)=0$, together with a small value (in magnitude) of the kinetic scalar at the 
onset of matter domination.\footnote{According to our conventions the kinetic scalar is negative definite on a cosmological background.} This can be easily 
guaranteed if we impose
\begin{equation}\label{eq.inequality.motion}
 \textrm{sign}(\alpha_B)=-\textrm{sign}(\alpha_K),
\end{equation}
as we explain below. 
The main plot can be summarized as follows: During inflation $1+2\dot{H}/3H^2= 1$, and the dynamical equation~(\ref{eq.dynamics0}) together with a 
different sign in the values of the braiding and kineticity functions can naturally explain the fall of the magnitude of the kinetic term up to the regime where 
$X= 0$, no matter its ``initial'' value at the big bang. During the radiation era, however, $1+2\dot{H}/3H^2=-1/3$, and the kinetic term will grow in magnitude from 
its small value at reheating up to the point where it remains frozen during matter domination, as we previously argued. Finally, the dark energy overtakes the other 
components and the kinetic term starts its way back to the region where $X=0$.  

As a consequence, the behavior of the scalar degree of freedom is mainly determined by the properties of $\bar{G}_4$ and $F_4$ close to the points $X=0$, and 
$1-M_*^2(1+\alpha_B)=\Omega_{\phi}$, where the last expression defines only implicitly the value of the kinetic term during the matter era. The particular details 
of $\bar{G}_4$ and $F_4$ between these two points are only relevant during the transitions from radiation to matter, and from matter to dark energy, domination, 
but they will not affect the behavior of the universe for most of its history. 

[Notice that there is also the possibility that the kinetic term gets stuck before matter-radiation equality if it comes across a point where the braiding coefficient 
$\alpha_B$ vanishes. As we have previously argued in Section~\ref{sec.DM}, the scalar mode tracks the dominant matter component in this case, hence contributing as a relativistic species 
during a period of time in the radiation era. Since the kinetic term is frozen, the dimensionless density parameter associated to the scalar mode will also remain constant. 
As a result of this, and if the density parameter is of order one during matter domination (as it should be to reproduce dark matter observations), it will be so also 
at the end of the radiation era. If the evolution led the kinetic term to a point where $\alpha_B=0$ before big-bang nucleosynthesis, the model would be ruled out 
by current constraints on the effective number of neutrino species at that time~\cite{Cyburt:2015mya}. But even if that is not the case, the presence of 
extra relativistic species will affect the time of matter-radiation equality, which is also well constrained by large scale structure and cosmic microwave background 
observations~\cite{Ade:2015xua}. This makes it very unlikely that the braiding function could vanish during matter domination, and this will be of some interest when 
exploring later the cosmological perturbations.]

This completes our qualitative description of the homogeneous background. However, in order to have a sensible cold dark matter mimicker, we need to prove the 
logarithmic and linear growth of scalar perturbations during the radiation and matter eras, respectively. In this paper we concentrate on the second of these 
periods, and conclude that it is no longer possible to reproduce the standard cold dark matter evolution. At this point the reader may refer to 
Appendix~\ref{app.perturbations}, where the details of linear order perturbations in presence of matter around an arbitrary homogeneous and isotropic universe are 
given for a general GLPV model. 

As an overview for the time-constrained reader, in the unitary gauge the scalar sector of the perturbations are governed by the following action,
\begin{eqnarray}\label{scalar.pert3}
 S^{(2)}_{\textrm{scalar}}=\frac{1}{2}\int dtd^3x a^3\left[\mathcal{A}_{ij}\dot{q}_i\dot{q}_j-a^{-2}\mathcal{C}_{ij}\partial q_i\partial q_j \right.\nonumber \\ 
 \left.-H\mathcal{B}_{ij}\dot{q}_i q_j-H^2\mathcal{D}_{ij}q_iq_j\right], 
\end{eqnarray}
where, as in Eq.~(\ref{eq.actionscalargravity}), we have integrated out the Hamiltonian and momentum constraints. The variables $q_i=(\zeta,v)$ codify the curvature 
perturbation, and total matter velocity perturbation, respectively, and the matrices $\mathcal{A}_{ij}$, $\mathcal{C}_{ij}$, $\mathcal{B}_{ij}$ and $\mathcal{D}_{ij}$ are 
defined in Eqs.~(\ref{matrices.AB}) and~(\ref{matrices.CD}). Varying the previous action with respect to 
$q_i$ yields the following equations of motion: 
\begin{equation}\label{eq.motion1}
 \ddot{q}_i + 3H(\delta_{ij}+\mathcal{E}_{ij})\dot{q}_j+H^2\mathcal{F}_{ij}q_j=0.
 \end{equation}
Notice that the matrices $\mathcal{E}_{ij}$ and $\mathcal{F}_{ij}$ contain all the dynamical information to evolve linear perturbations, and for a general GLPV 
model they are given in Eqs.~(\ref{matrix.E}) and~(\ref{matrix.F}). 

During matter domination, a cold dark matter component must satisfy $\mathcal{E}_{ij}=\mathcal{F}_{ij}=0$; see the discussion in Appendix~\ref{fiducialDM}. 
Since the kinetic scalar is frozen during the matter era, we can easily conclude that the time varying components of these matrices vanish, i.e. $\dot{\mathcal{A}}_{ij}=\dot{\mathcal{B}}_{ij}=0$. 
In order to guarantee a vanishing $\mathcal{E}_{ij}$, we need to impose the remaining part of Eq.~(\ref{matrix.E}) to be zero, which translates in the product $\mathcal{A}_{ik}^{-1}\mathcal{B}_{[kj]}=0$. 
In presence of matter, $\alpha_B\neq -1$, and it is possible to get convinced that this is equivalent to setting the only nonzero component of $\mathcal{B}_{ij}$, 
that we call $b$ in Eq.~(\ref{matrices.CD}), to zero. 

On the other hand, the two matrices $\mathcal{D}_{ij}$ and $\mathcal{C}_{ij}$ in Eq.~(\ref{matrix.F}) are linearly independent, so to further guarantee a 
vanishing $\mathcal{F}_{ij}$ we also need to demand that the two contributions $\mathcal{A}_{ik}^{-1}\mathcal{D}_{kj}$ and $\mathcal{A}_{ik}^{-1}\mathcal{C}_{kj}$ 
vanish independently. In a universe where $\alpha_B\neq -1$, this can only be satisfied if the single component, named $d$ in Eq.~(\ref{matrices.CD}), of 
$\mathcal{D}_{ij}$, and the speed of sound $c_s^2$ associated to the modified gravity terms in $\mathcal{C}_{ij}$, vanish.\footnote{If the matrix 
$\mathcal{A}_{ik}^{-1}\mathcal{C}_{kj}$ vanishes, their eigenvalues $\lambda$ must also be zero. They are determined in terms of the quadratic equation 
$\det(\lambda\delta_{ij}-\mathcal{A}_{ik}^{-1}\mathcal{C}_{kj})=0$, which can be written in the form $\det(\mathcal{A}_{ik}^{-1}) \det(\lambda \mathcal{A}_{kj}-\mathcal{C}_{kj})=0$. If both matter and gravity
modes propagate, then $\det(\mathcal{A}_{ik}^{-1})\neq 0$, and the eigenvalues $\lambda$ are nothing but their associated squared speeds of sound $c_m^2$ and $c_s^2$, respectively, see Appendix~\ref{app.perturbations}. 
The former one naturally vanishes in a universe dominated by nonrelativistic matter, $c_m^2=0$, so one only needs to impose $c_s^2=0$.}

The background functions $c_s^2$, $b$ and $d$ have been computed in Appendix~\ref{app.perturbations}, and their expressions in terms of the BS parameters 
can be found in Eqs.~(\ref{eq.b}), (\ref{eq.d}), and (\ref{eq.speeds}). After some algebra, and for the case of a universe
dominated by nonrelativistic particles, they can be reexpressed in the more convenient form
\begin{eqnarray}
c_s^2 &=& \left(\frac{1+\alpha_B}{\alpha_K+6\alpha_B^2}\right) \left[-2\alpha_B(1+\alpha_T)-2\alpha_T-\alpha_H\right],\;\;\label{eq.parameters.cs2} \\
b &=& 3\left(\frac{1+\alpha_B}{\alpha_K+6\alpha_B^2}\right)\left[\alpha_K-6\alpha_B\right], \label{eq.parameters.b}\\
d &=& 9(1+\alpha_B)\left(\frac{1+\alpha_B}{\alpha_K+6\alpha_B^2}\right)\left[3+6\alpha_B-\alpha_K/2\right].\label{eq.parameters.d}
\end{eqnarray}
Since $\alpha_B\neq -1$ in the presence of matter, and according to Eq.~(\ref{eq.dimensionless}), the only possible way in which $1+\alpha_B\to 0$ (keeping a 
reasonable value of $\Omega_{\phi}$), is that $M_*^2$ tends to infinity. If one naively extrapolates the constraints on the dark matter sound speed from 
Ref.~\cite{Kunz:2016yqy}, $c_{s(\textrm{DM})}^2\lesssim 10^{-10.7}$, the value of the cosmological strength of gravity must be at least ten billions larger than 
the inferred from local observations. Among other things, a larger value of $M_*^2$ would drastically increase the growth rate of structure formation, in clear 
disagreement with observations~\cite{Macaulay:2013swa}. This can be seen from the Newtonian potential evolution equations, given by e.g. Eq.~(4.1) in Ref.~\cite{Bellini:2014fua}, 
where the matter source term depends only on $M_*^2$, and not on the $(1+\alpha_B)$ factor that leaves the product $M_*^2(1+\alpha_B)$ of order one. 
Furthermore, even though $c_{s}^2=0$ if $\alpha_K+6\alpha_B^2\to\infty$, at least one of the 
expressions for $b$ or $d$ would remain of order one in that case. Moreover, having large values of the BS parameters might suggest the breakdown of the perturbative
EFT construction, leading to large higher order corrections. Therefore, we can safely conclude that in order to make $c_s^2=b=d=0$, 
the square brackets of Eqs.~(\ref{eq.parameters.cs2}), (\ref{eq.parameters.b}), and~(\ref{eq.parameters.d}) must vanish simultaneously. 

Apparently, there are many different ways in which the square bracket of $c_s^2$ can be zero, but according to the recent multimessenger observation, we must set 
$\alpha_T=0$ in Eq.~(\ref{eq.parameters.cs2}), reducing the number of possible choices. Contrary to the case of Section~\ref{sec.EdS} (where there was nothing but 
gravity and the kinetic scalar, and then also the tensor speed excess, was stucked at a fixed point during the evolution), now we need to guarantee $c_g^2=1$ not only at 
matter domination, but also during the transition to the dark energy era, when the kinetic term is not frozen but runs. According to the expressions in 
Appendix~\ref{functions}, this condition relates the two arbitrary functions of the Lagrangian~(\ref{eq:Action}), $F_4=2\bar{G}_{4X}/X$, and enforces 
$\alpha_H=- \alpha_B$. The square bracket of Eq.~(\ref{eq.parameters.cs2}) reduces to $-\alpha_B$ in this case, so the braiding coefficient must vanish during the 
matter era. This is however very unlikely, as we previously argued in this section. But even if one forgets those arguments and sets $\alpha_B=0$ into the bracket of 
Eq.~(\ref{eq.parameters.d}), the condition $d=0$ enforces $\alpha_K=6$. All the BS coefficients apart from the cosmological strength of gravity have been 
already determined in this model, and from Eq.~(\ref{eq.parameters.b}) we can read $b=6$. This induces a gravitational coupling between the perturbations
in the different components that is not present for a standard cold dark matter component.

In the standard cosmological scenario, dark matter perturbations dominate at matter-radiation equality. If one neglects the subleading 
contributions in Eq.~(\ref{eq.motion1}), and solves for the curvature perturbations, the familiar solution in Eq.~(\ref{eq.ev.perts}) is recovered. 
However, baryons are affected in a nontrivial way by the presence of dark matter in this model. Even if a detailed numerical analysis (using, e.g.,
{\tt hi-class}~\cite{Zumalacarregui:2016pph} or {\tt EFTCAMB}~\cite{Hu:2013twa}), is necessary in order to give some precise numbers, it seems very unlikely 
that an order one modification in any of the parameters of the dynamical equations could be consistent with the precision of current cosmological observations.

\section{Discussion}~\label{sec.discussion}

In this paper, we have explored some interesting cosmological consequences of the GLPV scalar-tensor theories~\cite{Gleyzes:2014dya,Gleyzes:2014qga}, where the scalar
degree of freedom plays the role of dark matter. These models generalize Horndeski gravity at the expense of introducing higher order equations of motion, but in 
such a way that no Ostrogradski instability propagates. In particular, we have analyzed a general sector of the theory that is invariant under parity and shift 
transformations in the scalar sector, and where the dynamics of the scalar mode is mainly determined by the nonminimally coupled terms. Other proposals with a similar spirit 
have been recently studied in the literature~\cite{Bettoni:2013zma,Rinaldi:2016oqp,Koutsoumbas:2017fxp}, but we can consider them as particular realizations of this broader scenario.

In the absence of matter, we have proved that the scalar mode can naturally drive the expansion of a homogeneous Einstein-de Sitter background, and at the same time 
clumps like standard cold dark matter particles. This behavior is independent of the particular expressions taken for the arbitrary functions $\bar{G}_4(X)$ and 
$F_4(X)$ in the Lagrangian, the only condition is that the braiding coefficient must be fixed to $\alpha_B=-1$. With an extra mild assumption on the 
parameters of the model necessary to guarantee that the tensor speed excesses vanishes, $\alpha_T=0$~\cite{Ezquiaga:2017ekz,Creminelli:2017sry,Baker:2017hug,Sakstein:2017xjx,Amendola:2017orw}, 
this subsector of the GLPV theory can be made consistent with the almost coincident observation of the gravitational wave signal GW170817 and its gamma ray counterpart GRB170817A. 

When standard model particles come into play, a nontrivial coupling between the scalar mode and matter modifies the previous picture, and a more elaborated analysis 
becomes necessary. At the level of the homogeneous background, the scalar degree of freedom still contributes as nonrelativistic particles during the matter era. Moreover, it 
is always possible to fit the two functions $\bar{G}_4(X)$ and $F_4(X)$ in such a way that the cosmological evolution matches the $\Lambda$CDM one at a desired 
accuracy. However, we proved that there is not any possible way to make that the combination of the extra mode and baryons evolve like in the standard cold dark 
matter scenario, even at the linear order in perturbations. 

To conclude, even if we leave the possibility of recovering a successful gravitationally coupled mode 
with some of the ingredients that we have elaborated in this paper open, a more sophisticated model seems required.  
If this model exists, then one should contrast it with linear observations, such as those of the cosmic microwave background radiation, and nonlinear evolution, such as 
large scale structure formation and galaxy dynamics. In this sense, it would be very interesting to identify which terms, if any, of the original effective 
Lagrangian are the responsible of the phenomenological correlations between dark matter and baryons that seems to emerge at small scales.
This is, however, beyond the scope of this paper.

\acknowledgments

F.F. acknowledges a CONACyT predoctoral grant. This  work was partially supported by CONACyT-Mexico under Grants No.~182445, No.~179208, No.~167335, and Fronteras de 
la Ciencia 281, by SEP-23-005 through Grant No. 18134, and also by DAIP-UG.

\appendix

\section{Bellini and Sawicki's parametrization}\label{functions}

For some parts of this paper we find convenient to use the Bellini and Sawicki's (BS) parametrization of  Horndeski gravity~\cite{Bellini:2014fua}, extended to GLPV 
models~\cite{Gleyzes:2014qga}, rather than the original functions in the Lagrangian~(\ref{eq:Hornd-action}). This parametrization consists on 5 
background functions, $M_*^2$, $\alpha_B$, $\alpha_K$, $\alpha_T$, and $\alpha_H$, describing the behavior of cosmological linear perturbations and that 
we summarize in this Appendix. For a GLPV model with $G_2=G_3=G_{4\phi}=F_{4\phi}=G_5=F_5=0$, such as the one in Eq.~(\ref{eq:Action}), these functions are given by
\begin{widetext}
\begin{eqnarray}
 M_*^2 &=& 1+2\bar{G}_4-4X\bar{G}_{4X}+2X^2F_4,\\
\alpha_B &=& -\frac{4}{M_*^2}\left[X\bar{G}_{4X}+2X^2\bar{G}_{4XX}-X^2(2F_4+XF_{4X})\right],\\
\alpha_K &=& \frac{12}{M_*^2}\left[X\bar{G}_{4X}+8X^2\bar{G}_{4XX}+4X^3\bar{G}_{4XXX}-X^2(6F_4+9XF_{4X}+2X^2F_{4XX})\right],\\
\alpha_T &=& \frac{2}{M_*^2}\left[2X\bar{G}_{4X}-X^2F_4\right],\label{eq.alphaT}\\
\alpha_H &=& -\frac{2}{M_*^2}X^2 F_{4}.
\end{eqnarray}
$M_*^2$ is the cosmological strength of gravity, $\alpha_B$ and $\alpha_K$ denotes the braiding and the kineticity functions, respectively, and $\alpha_T$ is the tensor speed 
excess. Only if the theory is beyond Hordeski $\alpha_H\neq 0$. Note that we are using the convention of Ref.~\cite{Gleyzes:2014qga}, where there is a $-1/2$ factor 
of difference with respect to the original parametrization~\cite{Bellini:2014fua}. Therefore, our definition of the braiding coefficient is $\alpha_{B}^{\textrm{here}}=-\alpha_{B}^{\textrm{there}}/2$.

\section{Linear perturbations}\label{app.perturbations}

In this Appendix we review cosmological linear order perturbation theory in the light of the GLPV gravity.
Although most of the expressions have been reported somewhere else, see, e.g., Refs.~\cite{Gleyzes:2014dya,Gleyzes:2014qga,DeFelice:2015isa}, others are new and necessary 
for the purposes of this paper.
To proceed, we expand the action in Eq.~(\ref{eq:Hornd-action}) to second order in perturbations for the case of a homogeneous and isotropic spatially 
flat universe that contains only standard model particles. For simplicity, we assume that this component can be described in terms of a perfect fluid with constant 
barotropic index, $p_{m}=w \rho_{m}$, and approximate $w=1/3$ ($w=0$) during the radiation (matter) era. Note that we are not imposing any restriction on the 
Lagrangian functions $G_i$ and $F_i$, and all the expressions below apply for general Horndeski and GLPV models, even though in this paper we are mainly concerned 
with the case of $G_2=G_3=G_{4\phi}=F_{4,\phi}=G_5=F_5=0$. 

Under this construction, the tensor sector of general relativity is modified to
\begin{equation}\label{eq.actiontensor}
S^{(2)}_{\textrm{tensor}}=\frac{M_*^2}{8}
\int dtd^3x a^3\left[\dot{\gamma}_{ij}^2-(1+\alpha_T)a^{-2}(\partial\gamma_{ij})^2\right],
\end{equation}
where $\gamma_{ij}$ is the transverse and traceless perturbation to the spatial metric. In order to prevent the appearance of ghosts we need a positive 
definite cosmological strength of gravity, $M_*^2>0$, whereas the absence of gradient instabilities imposes $1+\alpha_T\ge 0$ on the tensor speed excess.

Apart from tensor perturbations, an additional scalar mode to the matter fluid also propagates in the gravitational sector of this theory. If vector sources are present, they usually dilute with cosmological 
expansion, hence we do not consider them here. 
The scalar modes, on the contrary, play a central role in our description of dark matter. In the unitary gauge, and integrating out the Hamiltonian 
and momentum constraints, the scalar propagating degrees of freedom are described by the following action:
\begin{equation}\label{scalar.pert2}
 S^{(2)}_{\textrm{scalar}}=\frac{1}{2}\int dtd^3x a^3\left[\mathcal{A}_{ij}\dot{q}_i\dot{q}_j-a^{-2}\mathcal{C}_{ij}\partial q_i\partial q_j-H\mathcal{B}_{ij}\dot{q}_i q_j-H^2\mathcal{D}_{ij}q_iq_j\right],
\end{equation}
where $q_i=(\zeta,v)$ are the curvature perturbation, $^{(3)}R=-4a^{-2}\partial^2\zeta$, and the total matter covariant velocity perturbation (normalized to the 
Hubble rate), $\delta ụ_i=H^{-1}\partial_i v$.
The matrices $\mathcal{A}_{ij}$, $\mathcal{B}_{ij}$, $\mathcal{C}_{ij}$, and $\mathcal{D}_{ij}$ in Eq.~(\ref{scalar.pert2}) can be expressed in terms of the BS parametrization as 
\begin{align}\label{matrices.AB}
 &\mathcal{A}_{ij} =\left( \begin{array}{cc}
         Q_s+\frac{Q_m}{(1+\alpha_B)^{2}} & \frac{Q_m}{(1+\alpha_B)} \\
         \frac{Q_m}{(1+\alpha_B)} & Q_m
        \end{array}\right),& 
 &\mathcal{C}_{ij} = \left( \begin{array}{cc}
         Q_s(c_s^0)^2 & \frac{Q_m}{(1+\alpha_B)}(1+\alpha_H)(c_m^0)^2 \\
         \frac{Q_m}{(1+\alpha_B)}(1+\alpha_H)(c_m^0)^2 & Q_m(c_m^0)^2
        \end{array}\right),\\[10pt]
 &\mathcal{B}_{ij}  = Q_s b\times\delta_{i1}\delta_{j2}, & 
 &\mathcal{D}_{ij} =  Q_s d\times\delta_{i2}\delta_{j2}, \label{matrices.CD}
\end{align}
%
where $Q_s$, $Q_m$ are the kinetic coefficients,
\begin{equation}\label{eq.kinetic}
Q_s=\frac{M_*^2(\alpha_K+6\alpha_B^2)}{(1+\alpha_B)^2},\quad
Q_m=\frac{\rho_m(1+w)}{H^2w},
\end{equation}
and $(c_s^0)^2$, and $(c_m^0)^2$ the squared sound speeds, 
\begin{equation}\label{eq.speed}
(c_s^0)^2=\frac{1}{Q_s}\left[\frac{2}{a}\frac{d}{dt}\left(\frac{aM_*^2(1+\alpha_H)}{H(1+\alpha_B)}\right)-2M_*^2(1+\alpha_T)\right],\quad 
(c_m^0)^2=w,
\end{equation}
associated to the isolated gravity and matter theories, respectively. Moreover, the only nonvanishing components of the matrices $\mathcal{B}_{ij}$ and $\mathcal{D}_{ij}$ are given by
\begin{eqnarray}
 b &=& \frac{Q_m}{(1+\alpha_B)^2Q_s}\left[(\alpha_K-6\alpha_B)(c_m^0)^2+2(1+\alpha_B)\frac{\dot{H}}{H^2}+\frac{Q_m(c_m^0)^2}{M_*^2}\right],\label{eq.b}\\
 d &=& \frac{Q_m}{(1+\alpha_B)^2Q_s}\left\lbrace\frac{Q_m(c_m^0)^2}{2M_*^2}\left[\left(3+6\alpha_B-\frac{1}{2}\alpha_K\right)(c_m^0)^2
 \right.\right.\nonumber\\
&&\left.\left.
+(1+\alpha_B)\left(3+\alpha_M+\frac{\dot{H}}{H^2}\right)-
\frac{Q_m(c_m^0)^2}{2M_*^2}+\frac{\dot{\alpha}_B}{H}\right]-
(1+\alpha_B)\frac{1}{H}\frac{d}{dt}\left(\frac{\dot{H}}{H^2}\right)\right\rbrace.\label{eq.d}
\end{eqnarray}
These last terms are an effective ``friction" and ``mass" factors that couple the nonderivative terms in the Lagrangian~(\ref{scalar.pert2}).

In order to avoid ghostlike instabilities in the scalar sector we need to demand that the determinants of the principal sub-matrices of the kinetic term 
$\mathcal{A}_{ij}$ are all positive definite, which translates into $Q_s>0$ and $Q_m>0$, the former condition being equivalent to $\alpha_K+6\alpha_B^2>0$. The dispersion relations of the propagating modes are 
obtained from the zeros of the quartic polynomial $\textrm{det}[\mathcal{A}_{ij}\omega^2-\mathcal{C}_{ij}k^2]=0$. In a general case, the two scalar modes, namely the gravitational 
and the matter one, are mixed by nontrivial kinetic and gradient couplings, and the final expressions for the dispersion relations are not very illuminating. 
However, for the simpler case of nonrelativistic matter, $(c_s^0)^2=0$, they reduce to
\begin{equation}\label{eq.speeds}
c_s^2=(c_s^0)^2-\frac{Q_m (1+2\alpha_H)}{Q_s(1+\alpha_B)^2}(c_m^0)^2,\quad
c_m^2=(c_m^0)^2.
\end{equation}
Note that in this particular limit the matter sound speed is not affected by the gravitational sector, a result of some interest for the study of the linear 
perturbations during the matter domination era. Lastly and in order to avoid gradient instabilities we need to satisfy $c_s^2>0$ and $c_m^2>0$.

Varying the expression in Eq.~(\ref{scalar.pert2}) with respect to $q_i$, and moving to Fourier space, we obtain the following equations of motion:
\begin{equation}\label{eq.motion}
 \ddot{q}_i + 3H(\delta_{ij}+\mathcal{E}_{ij})\dot{q}_j+H^2\mathcal{F}_{ij}q_j=0,
\end{equation} 
where
\begin{eqnarray}
&& \mathcal{E}_{ij} = \frac{1}{3H}\mathcal{A}^{-1}_{ik}\left(\dot{\mathcal{A}}_{kj}-\mathcal{B}_{[kj]}\right),\label{matrix.E}\\
&& \mathcal{F}_{ij} = \mathcal{A}^{-1}_{ik}\left[-\frac{1}{2}\left(3+\frac{\dot{H}}{H^2}\right)\mathcal{B}_{kj}-\frac{1}{2}H^{-1}\dot{\mathcal{B}}_{kj}+\mathcal{D}_{kj}+H^{-2}C_{kj}a^{-2}k^2 \right],\label{matrix.F}
\end{eqnarray}
and, as usual, $\mathcal{B}_{[kj]}\equiv (\mathcal{B}_{kj}-\mathcal{B}_{jk})/2$.
Notice that the two matrices $\mathcal{E}_{ij}$ and $\mathcal{F}_{ij}$ contain all the information at linear order in perturbation theory.

\end{widetext}

\section{A fiducial dark matter model}\label{fiducialDM}

As long as the scales of interest are large enough, the velocity potential of an irrotational perfect fluid~\cite{Diez-Tejedor:2013nwa} with equation of state $w=0$ can properly describe the behavior
of a nonrelativistic dark matter degree of freedom.
At the effective level, we can model such a matter component in terms of a purely kinetic k-essence model with $G_2(X)=X^n$ and $G_3(X)=\bar{G}_4(X)=G_5(X)=0$, 
where $n=(1+w)/2w$. Note that we need to make $n\to\infty$ in order to get a vanishing barotropic index.

During matter domination, nonrelativistic standard model particles and dark matter contribute in a significant way to the energy density of the universe, and then both 
$\rho_{\textrm{CDM}}$ and $\rho_{m}$ remain of the same order, with $w=0$ for the two components. Introducing these assumptions into Eqs.~(\ref{eq.kinetic}), 
(\ref{eq.b}), (\ref{eq.d}), and~(\ref{eq.speeds}), we obtain
\begin{equation}
 Q_s=\frac{\rho_{\textrm{CDM}}(1+w)}{H^2w}, \quad c_s^2=w,\quad b=d=0.
\end{equation}
After some algebra, we can simplify the two matrices $\mathcal{E}_{ij}$ and $\mathcal{F}_{ij}$ in Eqs.~(\ref{matrix.E}) and~(\ref{matrix.F}) to get 
\begin{equation}\label{DMconditions}
 \mathcal{E}_{ij} = 0, \quad \mathcal{F}_{ij} = a^2w(k/H)^2\delta_{ij}.
\end{equation}
So far, we have just assumed that the  barotropic index $w$ is small, but nonzero in a mathematical sense. Note that even though the function $Q_s$ diverges as 
$w^{-1}$, all the coefficients in the equations of motion~(\ref{eq.motion}) remain finite, so one can obtain meaningful physical results from these expressions. 
If we take the limit $w\to 0$ at the end of the calculation, we can easily conclude that $\mathcal{E}_{ij}=\mathcal{F}_{ij}=0$ for a cold dark matter degree of freedom.

\end{document}